# Artificial Intelligence Assisted Power Grid Hardening in Response to Extreme Weather Events

**R. ESKANDARPOUR, A. KHODAEI**
University of Denver
USA

**A. PAASO, N. M. ABDULLAH**
ComEd
USA

**SUMMARY**

In this paper, an artificial intelligence based grid hardening model is proposed with the objective of improving power grid resilience in response to extreme weather events. At first, a machine learning model is proposed to predict the component states (either operational or outage) in response to the extreme event. Then, these predictions are fed into a hardening model, which determines strategic locations for placement of distributed generation (DG) units. In contrast to existing literature in hardening and resilience enhancement, this paper co-optimizes grid economic and resilience objectives by considering the intricate dependencies of the two. The numerical simulations on the standard IEEE 118-bus test system illustrate the merits and applicability of the proposed hardening model. The results indicate that the proposed hardening model through decentralized and distributed local energy resources can produce a more robust solution that can protect the system significantly against multiple component outages due to an extreme event.

**KEYWORDS**

Power grid resilience, machine learning, power grid hardening, extreme events.

Rozhin.Eskandarpour@du.edu

# 1. INTRODUCTION

Extreme weather events and natural disasters are the major cause of power outages in the United States, resulting in significant economic, social, and physical disruptions [1]. Various events have different characteristics and behaviour, however, the aftermath of all these events on the power grid is the loss of components and potential power outages. Among these events, hurricanes are explored in this paper not only because they cause the most widespread and long-lasting outages in the United States [2], but also because weather forecasting approaches that can predict a hurricane's arrival and characteristics (wind speed, hurricane type, duration etc.) are optimally advanced to determine the probable impact in a localized region [3]. Utilities and local governments are dealing with rising expectations of uninterrupted service from electricity consumers to effectively respond to the outcome of these catastrophic occurrences. Therefore, the topic of power grid resilience has received significant attention over the years as the number of these low-probability high-impact events increase.

With the purpose of improving the power grid resilience, electric utilities in the U.S. are spending billions of dollars on proactive and preventive responses such as grid hardening [4]. Grid hardening represents the physical and nonphysical improvement to the electricity infrastructure to make it less susceptible to adverse extreme events improving grid resilience and enabling the grid to withstand the impacts of extreme events with the least possible outages [5]. Physical hardening refers to installing new facilities and modifying the current grid topology. Nonphysical hardening options represent adjustments in consumption, generation, and power flow patterns. Current electric power grid hardening practices merely focus on the aspect of improving system resilience in responding efficiently to an extreme event.

In practice, multiple grid hardening options may be available for system planners. Finding the most suitable option is a challenging task as several factors are involved in the modelling, and furthermore mathematical approaches may not be able to fully capture the behaviour and aftermath of the events. Given the amount of data that exists on previous hurricanes and the complexity of the system, machine learning can be a viable approach to tackle this problem. Machine learning is an application of Artificial Intelligence (AI) that provides the system the ability to learn from historical data and to make predictions without being explicitly programmed. Machine learning approaches are utilized in a considerable number of research efforts in the power and energy sector, such as security assessment [6], load forecasting [7], distribution fault detection [8], and power outage duration prediction [9][10][11][12] [13] [14] [15] [16].

This paper proposes a computationally-efficient and economically-viable grid hardening model in response to ongoing challenges and urgent needs in designing more resilient power grids. First, the state of each component is predicted using a Support Vector Machine (SVM) which is trained on historical data with two features: the distance of the component from the center of the hurricane, and the category of the hurricane. Then, these predictions are fed to a hardening model, which takes grid resilience and economic needs into consideration. Different from existing literature in hardening and resilience enhancement, this paper identifies that investments targeted at resilience enhancement would indeed impact power grid resilience and economic operations. The proposed grid hardening model determines the economically optimal set of candidates to be deployed for enhancing system resilience under prevailing uncertainties, while ensuring an adequate and secure supply of forecasted loads under normal, contingency, and extreme conditions. The rest of the paper is organized as follows: Section 2 presents the model outline and formulation of the machine learning method used for outage prediction and the proposed hardening model. Section 3 presents simulation results on a test system. Finally, Section 4 concludes the paper.

## 2. COMPONENT OUTAGE PREDICTION AND GRID HARDENING MODEL

The outline of the proposed grid hardening model is depicted in Fig. 1. The problem is solved in three consecutive steps. In step 1, an SVM model is trained to classify the components into two states of damaged (on outage) and operational (in service) based on historical data. In step 2, the category and the path of an upcoming hurricane are forecasted which can be obtained from a weather forecasting channel. The category and path are used to identify the intensity of the hurricane and the potentially impacted regions, respectively. The speed of the hurricane and the distance of each power grid component from the center of the hurricane are used to predict the state of each component using the model trained in step 1. These predictions can subsequently help determine a set of suitable hardening candidates. Step 3 solves a grid hardening problem to ensure a secure supply of loads in response to the forecasted extreme event based on the predicted state of the components from step 2 and through strategic placement of utility-owned DGs. The proposed hardening model takes grid resilience and economic needs into consideration with the objective of minimizing the total system upgrade cost as well as system operation costs, subject to prevailing investment and operation constraints.

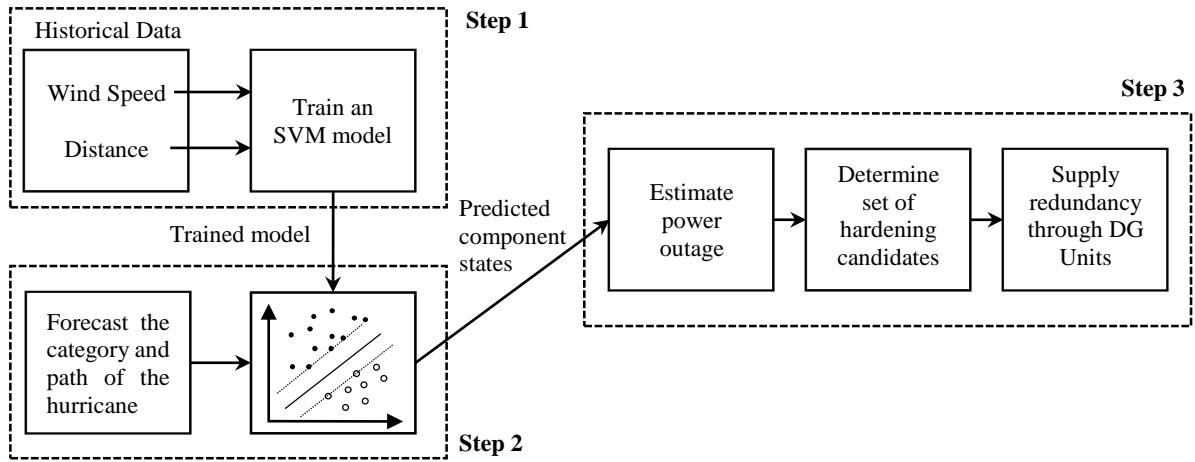

Fig. 1. Proposed grid hardening model

This paper focuses on physical hardening options, as resilience events are mainly triggered by outages and displacements of physical power grid facilities. Supply redundancy is considered as a valuable hardening approach. Supply redundancy decentralizes the electricity generation, thus instead of relying on large-scale power plants and bulk transmission network for power supply and delivery, a localized supply of power is utilized in certain regions to improve resilience. In this case, if power transfer and delivery from centralized generation is interrupted, a local supply of loads will be provided via available DGs.

### 2.1. Component Outage Prediction

The SVM is applied to classify the component states by defining a decision boundary into two classes (operational and outage) based on the wind speed and the distance of each component from the center of the hurricane. SVM is a binary classifier that separates training examples of one class from the other by defining a proper hyperplane. Assuming $x_\beta$ is the feature vector of each training example ($x_\beta \in R^D$) and $y_\beta$ is their corresponding class labels ($y_\beta = \pm 1$), the decision boundary hyperplane is defined as:

$$h_{w,g}(x_\beta) = \text{sgn}(w^T x_\beta + g) \tag{1}$$

where $w$ is the normal vector to the hyperplane separating training examples, $|g|/||w||$ is the perpendicular distance from the hyperplane to the origin, and *sgn* is the sign function, i.e., sgn($z$) = 1 if $z \geq 0$, and sgn($z$) = −1 otherwise. Function $h(x)$ is the output of the classifier with the aim of $h(x_\beta)=1$ if $y_\beta=+1$ and $h(x_\beta)=-1$ otherwise. The best hyperplane is defined as the hyperplane with the widest gap, known as the margin, and the training samples on the margin are called support vectors (shown in Fig. 2). The SVM determines the decision boundary hyperplane by solving the quadratic programming problem (2):

$$\min \frac{1}{2}\|w\|^2 + c\sum_{\beta=1}^{m} \varepsilon_\beta \tag{2}$$

s.t.

$$y_\beta(w^T x_\beta + g) \geq 1, -\varepsilon_\beta, \qquad \beta = 1,\ldots,m$$

$$\varepsilon_\beta \geq 0, \qquad \beta = 1,\ldots,m$$

where $c$ is a penalty parameter allowing separating nonlinear examples, and $\varepsilon_\beta$ is the regularization weight of the samples in the margin (support vectors). This quadratic programming problem can be solved by a Lagrange duality.

## 2.2. Grid hardening

The proposed grid hardening model minimizes the total investment cost of the grid hardening candidates as well as system operation costs, subject to prevailing investment and operation constraints. For reliability studies in power systems, it is common to use the *N*-1 criterion. The *N*-1 criterion simply states that the system needs to adequately and reliably supply loads in case of a single component outage at any given time. However, after an extreme event, it is anticipated that more than one component is affected and becomes unavailable. Hence, different contingency scenarios are considered in neighbouring locations along the hurricane path in which more than one component can be in outage state. Assuming *s* is the contingency scenario, the problem objective is defined as:

$$\min \sum_t \sum_i F_i(P_{it0}, I_{it}) + \sum_t \sum_s \sum_b v_b LC_{bts} + \sum_b IC_b P_b^{G,\max} \tag{3}$$

where $F_i(.)$ is the operation cost of unit $i$ in normal operation, $v$ is the value of lost load, $LC_{bts}$ is the amount load curtailment, and $IC_b$ is the investment cost associated with system upgrades by a DG unit with the capacity of $P_b^{G,max}$ at bus $b$. The value of lost load, $v$, is defined as the average cost that each type of customer, i.e., residential, commercial, or industrial, is willing to pay in order to avoid load interruptions [18]. Assuming $UX_{its}$ as the operation state of unit $i$ at time $t$ in scenario $s$ (1 when operating and 0 when on outage), and $UY_{lts}$ as the operation state of line *l* at time *t* in scenario *s* (1 when operating and 0 when on outage), the following operational constraints are defined:

$$\sum_{i \in B} P_{its} + \sum_{b \in B} P_{bts}^G + \sum_{l \in B} PL_{lts} + LC_{bts} = D_{bt} \qquad \forall b, \forall s, \forall t \qquad (4)$$

$$P_i^{\min} I_{it} UX_{its} \leq P_{its} \leq P_i^{\max} I_{it} UX_{its} \qquad \forall i, \forall s, \forall t \qquad (5)$$

$$\left| P_{it0} - P_{its} \right| \in \Delta_i \qquad \forall i, \forall s, \forall t \qquad (6)$$

$$\left| PL_{lts} - \frac{\sum_b a_{lb} \theta_{bts}}{x_l} \right| \leq M \left( 1 - UY_{lts} \right) \qquad \forall l, \forall s, \forall t \qquad (7)$$

$$-PL_l^{\max} UY_{lts} \leq PL_{lts} \leq PL_l^{\max} UY_{lts} \qquad \forall l, \forall s, \forall t \qquad (8)$$

$$0 \leq P_b^G \leq P_b^{G,\max} \qquad \forall b, \forall s, \forall t \qquad (9)$$

$$\sum_b IC_b P_b^{G,\max} \leq \Delta \qquad \forall b \qquad (10)$$

Constraint (4) represents nodal load balance. The load balance ensures that the total injected power to each bus from generation units, supply redundancies through DGs, and line flows is equal to the total consumed load at that bus. The load curtailment variable, *LC*, is added to the load balance equation to ensure a feasible solution when there is not sufficient generation to supply loads (due to component outages). Load curtailment is zero under normal operation conditions. Generation unit output power is limited by its capacity limit and is set to zero depending on its commitment and operation states (5). The change in a unit generation is further limited by the maximum permissible limit between normal and contingency scenarios (6). Transmission line capacity limits and power flow constraints are modeled by (7) and (8), respectively, in which the operation state is included to effectively model the line outages in contingency scenarios. $P^G_{bts}$ is the DG output power which is limited by its capacity limit and is set to zero depending on supply redundancy decision at bus *b* (9). Furthermore, the sum of the investment cost of all installed DGs in the system cannot exceed the available budget set by the system planner (10).

## 3. CASE STUDY

### 3.1. SVM performance

As historical data for the past extreme events at component level are limited, 600 samples are generated (300 samples of each component state, i.e., operational and outage) following a normal distribution function with a small Gaussian noise. The features are normalized to [0, 1] based on the maximum considered values of wind speed and distance. These samples are shown in Figure 2.

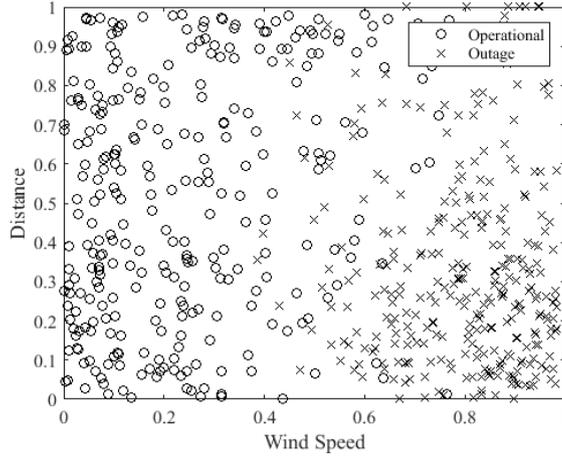

Fig. 2. Generated samples for each class (operational and outage)

To measure the performance of the proposed method, 20% of the samples (60 samples in outage state and 60 samples in operational state) are randomly selected to test/validate the SVM, and the remaining (80%) are used to train the model. Various penalty parameters ($c$=0.01, 0.1, 1, 10, 100) are trained. Among the trained SVM, $c$=1 represents the best overall classification accuracy on the validation set for the generated data, with overall classification accuracy of 92.5%. Table I shows the confusion matrix of classifying components into two classes of outage (having high probability of failure) and normal based on the distance to the center of the hurricane. As observed, the proposed method can classify the outage components from normal condition with a high accuracy.

Table I. Confusion matrix of classifying system components during extreme event (number of samples and percentage)

|  |  | Predicted | |
|---|---|---|---|
|  |  | Operational | Outage |
| Actual | Operational | 56 (93.33%) | 4 (6.66%) |
| Actual | Outage | 5 (8.33%) | 55 (91.66%) |

## 3.2. Improving grid resilience through the hardening model

The proposed hardening model is applied to the standard IEEE 118-bus test system. A hurricane is assumed to pass through three hypothetical paths as shown in Figure 3. The components in each path and its neighboring areas are classified into two categories of operational and outage according to the wind speed and the distance to the center of the hurricane, using the SVM model trained in the previous section. The trained model classified 48, 56, and 55 components as outage in paths 1, 2 and 3, respectively.

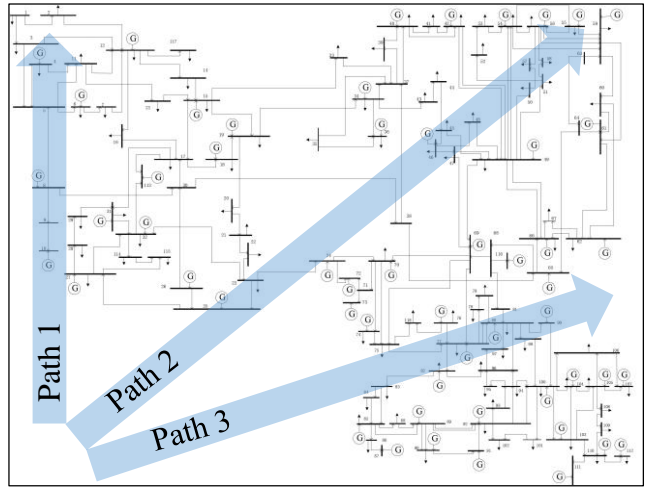

Fig. 3. IEEE 118-bus test system and the forecasted hurricane passing through three hypothetical paths

The proposed hardening model and the optimal scheduling problem is solved for one year (8760 hours). The value of lost load is considered $100/MWh at all buses. The investment cost associated with installing a DG unit (supply redundancy) at any given bus is assumed to be $50/MW. The following cases are studied:

**Case 1:** In this case, power grid scheduling is performed without hardening (supply redundancy). The optimal operation cost is obtained as $366,277,300. A total of 43338, 47143, and 44393 MWh load curtailment occurs in paths 1, 2, and 3, respectively. The average cost of unserved energy is calculated as $449,580,000.

**Case 2:** In this case, power grid scheduling is solved using the proposed hardening model. It is assumed that there is no constraint on investment budget. The annual optimal operation cost is obtained as $492,307,700. No load curtailment has occurred in this case, so the cost of unserved energy is zero and the system is secure against considered component outage scenarios. The proposed model advocates on hardening options at buses 33, 37, 39, 41, 42, 54, 59, and 80 to avoid load curtailments.

**Case 3:** This case discusses the effect of system hardening investment budget on the solution when all other parameters are kept unchanged. The results are summarized in Table II. As shown, the average unserved energy decreases by increasing the amount of budget.

Table II. Effect of investment budget on operation cost and load curtailment

| Budget | Load Curtailment (MWh) | | | Average Unserved Energy Cost |
| --- | --- | --- | --- | --- |
| | Path 1 | Path 2 | Path3 | |
| $0M | 43,338 | 47,143 | 44,393 | $449,580,000 |
| $1M | - | 22,341 | 3155 | $84,986,666 |
| $10M | - | 20,138 | 2,751 | $76,296,666 |
| $100M | - | 5294 | - | $17,646,666 |
| $126M | - | - | - | $0 |

As Table II suggests the relationship between the investment budget and average unserved energy cost reduction is not linear. For instance, the unserved energy cost reduced drastically ($364,593,334) with $1M investment, but to zero out the unserved energy cost (from $84,986,666 to zero), the system requires $125 M additional budget. The final decision is a trade-off between hardening budget and load curtailment reduction based on planner's discretion.

## 4. CONCLUSION

In this paper, an electric power grid hardening model was proposed through localized and decentralized supply of power in certain regions. At first, to identify hardening candidates, a machine learning model was proposed to predict the component states (either operational or outage) in response to a forecasted hurricane. Then, these predictions were fed to the proposed hardening model, which took resilience as well as the investment cost associated with system upgrades and decentralized supply of power into consideration. The numerical simulations on the standard IEEE 118-bus test system illustrated the merits and applicability of the proposed hardening model. The results indicated that the proposed hardening model can produce a robust solution that can protect the system against multiple component outages due to a hurricane.